\documentclass[]{article}
\usepackage{lmodern}
\usepackage{amssymb,amsmath}
\usepackage{ifxetex,ifluatex}
\usepackage{fixltx2e} 
\ifnum 0\ifxetex 1\fi\ifluatex 1\fi=0 
  \usepackage[T1]{fontenc}
  \usepackage[utf8]{inputenc}
\else 
  \ifxetex
    \usepackage{mathspec}
  \else
    \usepackage{fontspec}
  \fi
  \defaultfontfeatures{Ligatures=TeX,Scale=MatchLowercase}
\fi
\IfFileExists{upquote.sty}{\usepackage{upquote}}{}
\IfFileExists{microtype.sty}{%
\usepackage{microtype}
\UseMicrotypeSet[protrusion]{basicmath} 
}{}
\usepackage[margin=1in]{geometry}
\usepackage{hyperref}
\hypersetup{unicode=true,
            pdftitle={SDRcausal: an R package for causal inference based on sufficient dimension reduction},
            pdfauthor={Mohammad Ghasempour; Xavier de Luna},
            pdfborder={0 0 0},
            breaklinks=true}
\usepackage{color}
\usepackage{fancyvrb}

\DefineVerbatimEnvironment{Highlighting}{Verbatim}{commandchars=\\\{\}}
\usepackage{framed}
\definecolor{shadecolor}{RGB}{248,248,248}
\newenvironment{Shaded}{\begin{snugshade}}{\end{snugshade}}

\newcommand{\CommentTok}[1]{\textcolor[rgb]{0.56,0.35,0.01}{\textit{#1}}}

\newcommand{\DataTypeTok}[1]{\textcolor[rgb]{0.13,0.29,0.53}{#1}}
\newcommand{\DecValTok}[1]{\textcolor[rgb]{0.00,0.00,0.81}{#1}}

\newcommand{\KeywordTok}[1]{\textcolor[rgb]{0.13,0.29,0.53}{\textbf{#1}}}
\newcommand{\NormalTok}[1]{#1}
\newcommand{\OperatorTok}[1]{\textcolor[rgb]{0.81,0.36,0.00}{\textbf{#1}}}

\newcommand{\StringTok}[1]{\textcolor[rgb]{0.31,0.60,0.02}{#1}}

\usepackage{graphicx,grffile}
\makeatletter
\def\maxwidth{\ifdim\Gin@nat@width>\linewidth\linewidth\else\Gin@nat@width\fi}
\def\maxheight{\ifdim\Gin@nat@height>\textheight\textheight\else\Gin@nat@height\fi}
\makeatother
\setkeys{Gin}{width=\maxwidth,height=\maxheight,keepaspectratio}
\IfFileExists{parskip.sty}{%
\usepackage{parskip}
}{
\setlength{\parindent}{0pt}
\setlength{\parskip}{6pt plus 2pt minus 1pt}
}
\ifx\paragraph\undefined\else
\let\oldparagraph\paragraph
\renewcommand{\paragraph}[1]{\oldparagraph{#1}\mbox{}}
\fi
\ifx\subparagraph\undefined\else
\let\oldsubparagraph\subparagraph
\renewcommand{\subparagraph}[1]{\oldsubparagraph{#1}\mbox{}}
\fi

\let\rmarkdownfootnote\footnote%
\def\footnote{\protect\rmarkdownfootnote}

\usepackage{titling}

  \title{SDRcausal: an R package for causal inference based on sufficient
dimension reduction}
    \pretitle{\vspace{\droptitle}\centering\huge}
  \posttitle{\par}
    \author{Mohammad Ghasempour\footnote{Department of Statistics, USBE, Umeå
  University,
  \href{mailto:mohammad.ghasempour@umu.se}{\nolinkurl{mohammad.ghasempour@umu.se}}} \\ Xavier de Luna\footnote{Department of Statistics, USBE, Umeå University,
  \href{mailto:xavier.deluna@umu.se}{\nolinkurl{xavier.deluna@umu.se}}}}
    \preauthor{\centering\large\emph}
  \postauthor{\par}
      \predate{\centering\large\emph}
  \postdate{\par}
    \date{}

\usepackage{float}

\usepackage{bbm} \usepackage{xcolor} \usepackage{pdfpages}

\begin{document}
\maketitle


{
\setcounter{tocdepth}{3}
\tableofcontents
}
\hypertarget{abstract}{%
\section{Abstract}\label{abstract}}

SDRcausal is a package that implements sufficient dimension reduction methods for causal inference as proposed in Ghosh, Ma, and de Luna (2021). The package implements (augmented) inverse probability weighting and outcome regression (imputation) estimators of an average treatment effect (ATE) parameter. Nuisance models, both treatment assignment probability given the covariates (propensity score) and outcome regression models, are fitted by using semiparametric locally efficient dimension reduction estimators, thereby allowing for large sets of confounding covariates. Techniques including linear extrapolation, numerical differentiation, and truncation have been used to obtain a practicable implementation of the methods. Finding the suitable dimension reduction map (central mean subspace) requires solving an optimization problem, and several optimization algorithms are given as choices to the user. The package also provides estimators of the asymptotic variances of the causal effect estimators implemented. Plotting options are provided. The core of the methods are implemented in C language, and parallelization is allowed for. The user-friendly and freeware R language is used as interface. The package can be downloaded from Github repository: \url{https://github.com/stat4reg}. 
\begin{itemize}
	\item SDRcausal is an R-package to perform causal inference in situations with large
amount of potentially confounding covariates. 
	\item Semiparametric sufficient dimension reduction methods are used to deal with high
dimensionality. 
	\item Average causal effect of binary treatments can be
obtained using different semiparametric estimators.
\end{itemize}
\textbf{Keywords}: average treatment effect; central mean subspace;
double robust estimation; semiparametric inference

\begin{figure}[h]
  \includegraphics{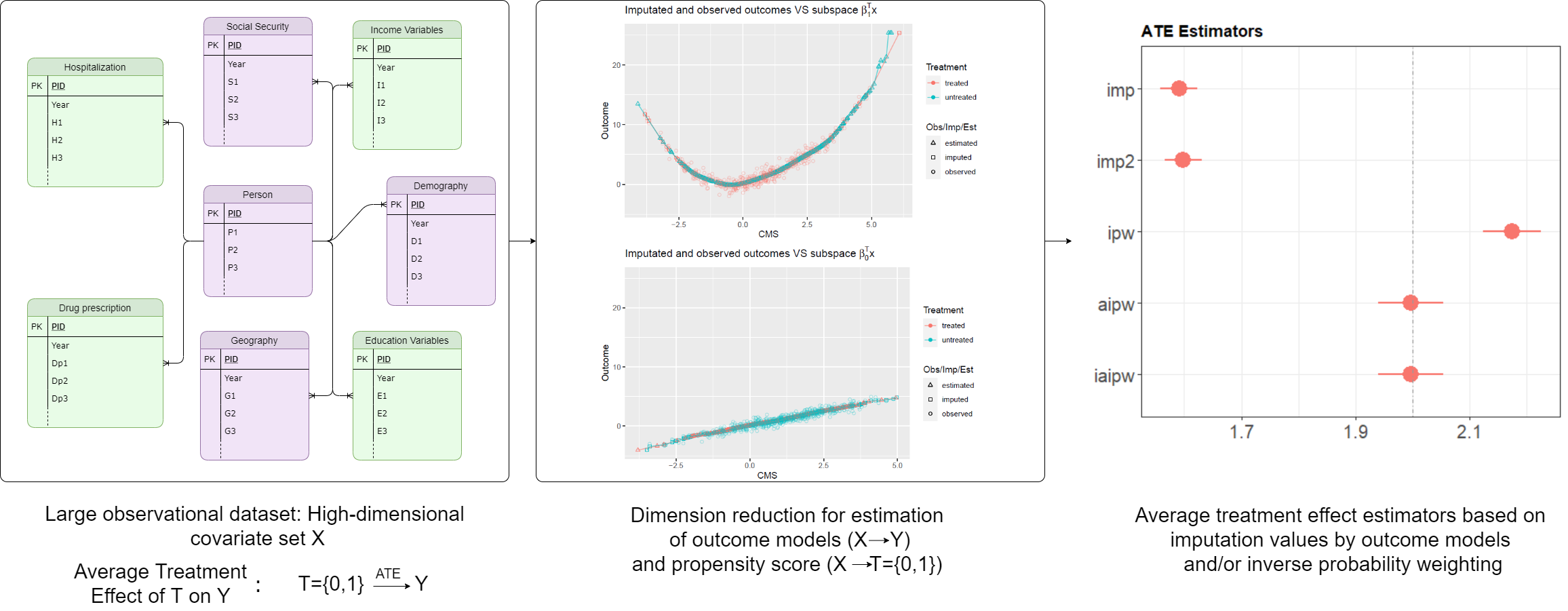}
  \caption{Graphical Abstract}
\end{figure}

\hypertarget{package-installation}{%
\section{Package installation}\label{package-installation}}

To install and load the package form GitHub, type the following command
in an R console:

\begin{Shaded}
\begin{Highlighting}[]
\KeywordTok{install.packages}\NormalTok{(}\StringTok{"devtools"}\NormalTok{)}
\KeywordTok{library}\NormalTok{(devtools) }
\KeywordTok{install_github}\NormalTok{(}\StringTok{"stat4reg/SDRcausal"}\NormalTok{)}
\KeywordTok{library}\NormalTok{(SDRcausal)}
\end{Highlighting}
\end{Shaded}

Mac users may also need to run the following command in a terminal
window before installing the package: \texttt{xcode-select\ -\/-install}

\hypertarget{assumptions-and-parameter-of-interest}{%
\section{Assumptions and parameter of
interest}\label{assumptions-and-parameter-of-interest}}

\hypertarget{average-treatment-effect}{%
\subsection{Average Treatment Effect}\label{average-treatment-effect}}

Let \(X_i \in\mathbb{R}^p\) be a set of covariates for the \(i\)th
individual in a sample of size \(n\). Assume there is a binary treatment
s.t. for each individual \(T_i=1\) or \(0\) if the \(i\)th individual
received treatment or not, respectively. Let \(Y_{1i}\) and \(Y_{0i}\)
be potential outcomes, i.e.~the outcomes under \(T_i = 1\) and
\(T_i = 0\) respectively. Only one potential outcome is observed for
each individual. We assume that \(Y_{i} = Y_{1i}T + Y_{0i}(1-T)\) is the
observed outcome. This package performs inference on the average causal
effect of treatment (or average treatment effect, ATE),
\[D = E(Y_1)-E(Y_0),\] using sufficient dimension reduction techniques.
This parameter is identified under the model given below.

\hypertarget{nuisance-models-and-dimension-reduction}{%
\subsection{Nuisance Models and Dimension
Reduction}\label{nuisance-models-and-dimension-reduction}}

Nuisance models are typically considered to identify the causal
parameter \(D\). These models are assumed to have signal belonging to
three different projections \(\alpha^T x\), \(\beta_0^T x\), and
\(\beta_1^T x\), i.e.~lower dimensional spaces. First, the propensity
score is modeled:

\[pr(T = 1 | X = x,Y_1,Y_0)=pr(T = 1 | X = x) = e^{\eta (\alpha ^T x)} / \{ 1 + e^{\eta (\alpha ^T x)} \},\]
where \(\eta (\cdot)\) is a smooth and bounded function. Boundedness
also guarantees that the propensity score is strictly in \((0, 1)\). The
parameter \(\alpha\) is a one or two dimensional projection matrix,
i.e.~with dimension \(p \times d\), \(d = 1 \ \text{or} \ 2\).

Furthermore, models for \(Y_1\) and \(Y_0\) given \(X = x\) are
specified as \[Y_1 = m_1(\beta^T_1 x) + \varepsilon_1 ,\] and
\[Y_0 = m_0(\beta^T_0 x) + \varepsilon_0 ,\] where
\(E(\varepsilon_1 | x)=E(\varepsilon_0 | x) = 0\). Here, \(m_1(\cdot)\)
and \(m_0(\cdot)\) are unknown functions, and \(\beta_1\) and
\(\beta_0\) are unknown projection matrices with dimension
\(p \times d, \ d=1,2\). These three models are estimated separately
using the random sample at hand. Then, by combining the models, fitted
in various ways, different estimators for the treatment effect
\(D = E(Y_1 - Y_0)\) are obtained.

\hypertarget{fitting-the-propensity-score-and-outcome-models}{%
\subsection{Fitting the Propensity Score and Outcome
Models}\label{fitting-the-propensity-score-and-outcome-models}}

The algorithms for fitting the outcome and propensity score models
through semiparametric dimension reduction are described in detail in
Ghosh, Ma, and de~Luna (2021). The goal of these algorithms is to find
\(\hat\alpha\), \(\hat\eta (\cdot)\), \(\hat\beta_t\), and
\(\hat m_t(\cdot)\), \(t=0,1,\) that satisfy the following conditions:

\begin{equation}
\sum_{i=1}^n t_i \{ y_{1i} -\hat m_1(\hat\beta^T_1x_i ) \}\hat m_1^\prime (\hat\beta_1^T x_i) \otimes \{ x_{Li} - \hat{E}(X_{Li} |\hat\beta_1^T x_i) \}  =0, \label{e1}
\end{equation}

\begin{equation} 
\sum_{i=1}^n t_i \{ y_{0i} -\hat m_0(\hat\beta^T_0x_i) \}\hat m_0^\prime (\hat\beta_0^T x_i) \otimes \{ x_{Li} - \hat{E}(X_{Li} |\hat\beta_0^T x_i) \}  =0, 
\label{e2}
\end{equation}

\begin{equation}
\sum_{i=1}^n \{x_{Li} - \hat{E}(X_{Li} |\hat \alpha^T x_i) \} \bigg[ t_i - \frac{e^{\hat\eta(\hat\alpha^T x_i)}}{1+ e^{\hat\eta(\hat\alpha^T x_i)}} \bigg] \hat\eta^\prime(\hat\alpha^T x_i)^T = 0, 
\label{e3}
\end{equation}

where \(X_{Li}\) is the subvector containing the lower \(p-d\)
components of \(X_{i}\), where \(X_{i}\) is the random vector of
covariates corresponding to \(i\)th individual.

The algorithms have two phases. In the first phase, the algorithms start
by finding naïve estimates of the functions \(\eta\) and \(m_t\),
\(t=0,1\). Initial guesses for \(\alpha\) and \(\beta_t\), \(t=0,1\),
matrices of dimension \(p\times d\), \(d=1,2\), are needed for these
naïve estimations. Then, the naïve estimates \(\hat{\eta}\) and
\(\hat{m_t}\) are in turn used to estimate \(\alpha\) and \(\beta_t\),
\(t=0,1\). The second phase consists of a repetition of the first part
of phase one by updating the initial estimates of \(\alpha\) and
\(\beta_t\), i.e., instead of initial guesses, the estimates of
\(\alpha\) and \(\beta_t\), obtained at the end of phase one, are used
to make new estimations for \(\eta\) and \(m_t\). Thus, first,
\(\hat{E}(X_{Li} | \alpha^T x_i)\) and
\(\hat{E}(X_{Li} | \beta^T_t x_i)\) are fitted based on initial guesses
of the projection parameters. Here, the Nadaraya-Watson kernel estimator
is used:
\[ \widehat{E}(X_L |\beta_t^T x) = \frac{\sum_{i=1}^n x_{Li} K_h (\beta_t^T x_i - \beta_t^T x)}{\sum_{i=1}^n K_h (\beta_t^T x_i - \beta_t^T x)},  \]
\[ \widehat{E}(X_L |\alpha^T x) = \frac{\sum_{i=1}^n x_{Li} K_h (\alpha^T x_i - \alpha^T x)}{\sum_{i=1}^n K_h (\alpha^T x_i - \alpha^T x)},  \]
where \(K_h (\alpha^T x_i - \alpha^T x)\) is such that
\(\frac{1}{h}K (\frac{\alpha^T x_i - \alpha^T x}{h})\), where different
kernel functions and bandwidth \(h\) can be used (see below).

The functions \(\eta\) and \(m_t\), \(t=0,1\), are then fitted using
local linear estimation. For the function \(m_t\):

\[
\widehat{m}_t(\beta_t^T x) =  A_{11} + A^T_{13} (A_{14} - A_{13}A^T_{13})^{-1}A_{13}A_{11},
\] where \begin{equation}
\begin{split}
A_{11} &= \frac{\sum_{i=1}^n t_iy_iK_h(\beta_t^T x_i - \beta_t^T x)}{\sum_{i=2}^n t_iK_h(\beta_t^T x_i - \beta_t^T x)}, \\
A_{13} &= \frac{\sum_{i=1}^n t_i(\beta_t^T x_i - \beta_t^T x)K_h(\beta_t^T x_i - \beta_t^T x)}{\sum_{i=2}^n t_iK_h(\beta_t^T x_i - \beta_t^T x)}, \\
A_{14} &= \frac{\sum_{i=1}^n t_i(\beta_t^T x_i - \beta_t^T x)(\beta_t^T x_i - \beta_t^T x)^TK_h(\beta_t^T x_i - \beta_t^T x)}{\sum_{i=2}^n t_iK_h(\beta_t^T x_i - \beta_t^T x)}.
\end{split}
\end{equation}

The following equations are solved for \(\eta\) and \(\eta^\prime\)
(deivative of \(\eta\)) for \(z = \alpha^Tx_1,...,\alpha^Tx_n\):

\begin{equation}
\begin{split}
&\sum_{i=1}^n \big[ t_i - \frac{e^{\eta(\alpha^Tx_i) + \eta^\prime(\alpha^Tx_i)^T(\alpha^T x_i - z)}}{1+ e^{\eta(\alpha^Tx_i) + \eta^\prime(\alpha^Tx_i)^T(\alpha^T x_i - z)}}  \big] K_h (\alpha^T x_i -z) = 0, \\
&\sum_{i=1}^n \big[ t_i - \frac{e^{\eta(\alpha^Tx_i) + \eta^\prime(\alpha^Tx_i)^T(\alpha^T x_i - z)}}{1+ e^{\eta(\alpha^Tx_i) + \eta^\prime(\alpha^Tx_i)^T(\alpha^T x_i - z)}}  \big] (\alpha^T x_i - z) K_h (\alpha^T x_i -z) = 0.
\end{split}
\end{equation}

The default kernel function in the package is
``Epanechnikov''(``parabolic''). Users can choose kernels and
bandwidths. The package support three kernels: ``Epanechnikov'',
``Quartic''(``biweight''), and ``Gaussian''.\\
The asymptotic properties of the estimators are derived in Ghosh, Ma,
and de~Luna (2021) based on conditions on the bandwidths; loosely a
range of proper bandwidths. In this framework the set of proper
bandwidths for \(K_h(\beta_t^T x_i - \beta_t^T x)\) is
\(h_c = c \ \text{sd}(\beta_t^Tx)n^{\frac{1}{5}}\), where \(c\) is an
arbitrary scale and \(\text{sd}(\beta_t^Tx)\) is the estimated standard
deviation of \(\beta_t^Tx\). (The same statement is valid for
\(\alpha^T x\).) The package uses these appropriate bandwidths based on
a scale \(c\) that is specified by the user.

\hypertarget{estimation-of-ate}{%
\section{Estimation of ATE}\label{estimation-of-ate}}

\hypertarget{a-data-generating-process}{%
\subsection{A Data Generating Process}\label{a-data-generating-process}}

The data generating process (Study 1 in Ghosh, Ma, and de~Luna (2021))
that we present in this section is used to illustrate the package
functions in the next sections. The seed that we used for the data
generation process is 48371. We set sample size to \(n = 1000\) and
covariate dimension \(p = 6\). Then, we generate the covariate vectors
\(X_i = (X_{1i}, . . ., X_{6i})^\intercal\) as:

\begin{itemize}
\item
  \(X_1 \sim N(1, 1)\) and \(X_2 \sim N(0, 1)\),
\item
  \(X_4 = 0.015 X_1 + u_1\), where \(u_1\) is uniformly distributed in
  \((-0.5, 0.5)\),
\item
  \(X_3 \sim \text{Bernoulli}(0.5 + 0.05X_2)\) and
  \(X_5 \sim \text{Bernoulli}(0.4 + 0.2X_4)\),
\item
  \(X_6 = 0.04X_2 + 0.15X_3 + 0.05X_4 + u_2\), where
  \(u_2 \sim N(0, 1)\).
\end{itemize}

In addition, we set \(\beta_1 = (1, -1, 1, -2, -1.5, 0.5)^T\),
\(\beta_0 = (1, 1, 0, 0, 0, 0)^T\) and
\(\alpha = (-0.27, 0.2, -0.15, 0.05, 0.15, -0.1)^T\).

We generate the response variables based on
\(Y_1 =0.7(\beta_1^T X)^2 + \sin(\beta_1^T X) + \varepsilon_1\) and
\(Y_0 = \beta_0^T X + \varepsilon_0\). Here \(\varepsilon_1\) and
\(\varepsilon_0\) are normally distributed with mean zero and variances
\(0.5\) and \(0.2\) respectively. Further, we let
\(\eta (\alpha^T x) = \alpha^T x\). Thus, the treatment indicator \(T\)
is generated from the logistic model
\(pr(T = 1|X) = \exp(\alpha^T x)/\{1 + \exp(\alpha^T x)\}\).

The generated dataset is stored in the package, and it is accessible by
using the following commands:

\begin{Shaded}
\begin{Highlighting}[]
\NormalTok{beta0 =}\StringTok{ }\NormalTok{SDRcausal}\OperatorTok{::}\NormalTok{beta0_guess}
\NormalTok{beta1 =}\StringTok{ }\NormalTok{SDRcausal}\OperatorTok{::}\NormalTok{beta1_guess}
\NormalTok{alpha =}\StringTok{ }\NormalTok{SDRcausal}\OperatorTok{::}\NormalTok{alpha_guess}
\NormalTok{covariates =}\StringTok{ }\NormalTok{SDRcausal}\OperatorTok{::}\NormalTok{covariates}
\NormalTok{y =}\StringTok{ }\NormalTok{SDRcausal}\OperatorTok{::}\NormalTok{outcomes}
\NormalTok{treated =}\StringTok{ }\NormalTok{SDRcausal}\OperatorTok{::}\NormalTok{treated}
\end{Highlighting}
\end{Shaded}

\hypertarget{imputation-through-outcome-regression-imp}{%
\subsection{Imputation through outcome regression
(IMP)}\label{imputation-through-outcome-regression-imp}}

\hypertarget{description}{%
\subsubsection{Description}\label{description}}

The first estimator is based on the outcome regression models, often
called an imputation approach (Ghosh, Ma, and de~Luna (2021), Section
3.1). Let

\[\widehat{E}_{\text{IMP}}(Y_1) = n^{-1} \sum_{i=1}^n \{t_iy_i + (1 - t_i)\widehat{m}_1(\hat{\beta}_1^T x_i) \},\]
and

\[\widehat{E}_{\text{IMP}}(Y_0) = n^{-1} \sum_{i=1}^n \{(1-t_i)y_i +  t_i\widehat{m}_0(\hat{\beta}_0^T x_i) \}.\]
Then the imputation estimator of \(D\) is:
\[\widehat{D}_{\text{IMP}} = \widehat{E}_{\text{IMP}}(Y_1) - \widehat{E}_{\text{IMP}}(Y_0). \]

Furthermore, there is an alternative imputation estimator that uses the
predicted values everywhere, even when the observed response values are
available, i.e.:

\[\widehat{E}_{\text{IMP2}}(Y_1) = n^{-1} \sum_{i=1}^n \widehat{m}_1(\hat{\beta}_1^T x_i),\]
and

\[\widehat{E}_{\text{IMP2}}(Y_0) = n^{-1} \sum_{i=1}^n \widehat{m}_0(\hat{\beta}_0^T x_i),\]
and the IMP2 estimator of \(D\) is:
\[\widehat{D}_{\text{IMP2}} = \widehat{E}_{\text{IMP2}}(Y_1) - \widehat{E}_{\text{IMP2}}(Y_0).\]

\hypertarget{example-and-code}{%
\subsubsection{Example and Code}\label{example-and-code}}

The imputation estimators are implemented in the function
\texttt{imp.ate} in the package. Initial guesses for \(\beta_1\) and
\(\beta_0\) are needed to call the function. Based on how good the
initial guesses are, users can choose different optimizers. If the
initial guess is close enough to the true values, choosing a local
optimizer is more efficient. Otherwise, if the initial guess is entirely
random, then using a global optimizer is probably a safer option. Two
different functions are available for optimization: \texttt{optim} and
\texttt{cobyla}. The function \texttt{optim} includes both global and
local optimizers. However, all the \texttt{optim} methods need the
derivability condition. If derivative-free optimization methods are
required, the \texttt{cobyla} method in the \texttt{nloptr} package is
recommended. By default \texttt{optim} is used, with the method
\texttt{Nelder-Mead}. To use another method, such as the global
optimization method \texttt{SANN}, just add \texttt{method\ =\ "SANN"}
to the \texttt{imp.ate} inputs. Other methods that can be used in
\texttt{optim} are: \texttt{BFGS}, \texttt{CG}, \texttt{L-BFGS-B}, and
\texttt{Brent}.

The default kernel function that is used to do the kernel regressions is
Epanechnikov, and it can be changed to Quartic or Gaussian by adding
\texttt{kernel\ =\ "QUARTIC"} or \texttt{kernel\ =\ "GAUSSIAN"}.

If \texttt{explicit\_bandwidth} is set as \texttt{FALSE}, bandwidths are
computed by the formula \(c \ \text{sd}{(\beta^Tx)} n^{\frac{1}{5}}\),
where c is determined by choosing values for \texttt{bwc\_dim\_red1},
\texttt{bwc\_impute1},\texttt{bwc\_dim\_red0}, and
\texttt{bwc\_impute0}. The values \texttt{bwc\_dim\_red1} and
\texttt{bwc\_dim\_red0} are used for all kernel regressions in the first
phase (finding a projection) of estimation of \(m_1(\beta_1^T x)\) and
\(m_0(\beta_0^T x)\), respectively. The values \texttt{bwc\_impute1} and
\texttt{bwc\_impute0} are used in the second phase (estimating
\(m_1(\cdot)\) and \(m_0(\cdot)\), respectively). If
\texttt{explicit\_bandwidth\ =\ TRUE} has been chosen, then the function
considers \texttt{bwc\_dim\_red1},
\texttt{bwc\_impute1},\texttt{bwc\_dim\_red0}, and \texttt{bwc\_impute0}
as the bandwidth values.

\begin{Shaded}
\begin{Highlighting}[]
\NormalTok{imp <-}\StringTok{ }\KeywordTok{imp.ate}\NormalTok{(}\DataTypeTok{x =}\NormalTok{ covariates,}
               \DataTypeTok{y =}\NormalTok{ y,}
               \DataTypeTok{treated =}\NormalTok{ treated,}
               \DataTypeTok{beta_guess1 =}\NormalTok{ beta1,}
               \DataTypeTok{beta_guess0 =}\NormalTok{ beta0,}\DataTypeTok{n_threads =} \DecValTok{2}\NormalTok{)}
\end{Highlighting}
\end{Shaded}

The list of all arguments is:

\begin{itemize}
\item
  \texttt{x} \newline Covariate matrix
\item
  \texttt{y} \newline Response vector
\item
  \texttt{treated1} \newline A binary vector indicating treatment status
\item
  \texttt{beta\_guess1} \newline Initial guess for \(\beta_1\)
\item
  \texttt{beta\_guess0} \newline Initial guess for \(\beta_0\)
\item
  \texttt{solver}~\\
  Specifies which solver is to be used. Current options are
  \texttt{"optim"} and \texttt{"cobyla"} (from \texttt{nloptr} package).
  The default value is \texttt{"optim"}.
\item
  \texttt{kernel}~\\
  Specifies which kernel function is to be used, current options are:
  \texttt{"EPAN"}(default), \texttt{"QUARTIC"}, and \texttt{"GAUSSIAN"}.
\item
  \texttt{explicit\_bandwidth}~\\
  Specifies if \texttt{bandwidth\_scale} will be used as the bandwidth
  or if it will be calculated as \texttt{bandwidth\_scale} \(\times\)
  sd(\(\beta^T x\)) \(\times\) \(n^{(1/5)}\). The default value is
  \texttt{FALSE}.
\item
  \texttt{recalc\_bandwidth}~\\
  Specifies whether the bandwidth should be recalculated after the first
  stage (the estimations of dimension reduction step). If
  \texttt{explicit\_bandwidth} = \texttt{TRUE},
  \texttt{recalc\_bandwidth} is not used, but if
  \texttt{explicit\_bandwidth} = \texttt{FALSE}, then if
  \texttt{recalc\_bandwidth} = \texttt{TRUE}, bandwidths are
  recalculated at the beginning of the second step based on
  \texttt{bwc\_impute0} and \texttt{bwc\_impute1}. If
  \texttt{recalc\_bandwidth} = \texttt{FALSE}(default), the first step
  bandwidths are used.
\item
  \texttt{bwc\_dim\_red1}~\\
  Scaling of calculated bandwidth, or if
  \texttt{explicit\_bandwidth\ =\ TRUE} used as the bandwidth. It is
  used in the dimension reduction step for \(\hat{m}_1(\beta_1^T x)\).
  The default value is \texttt{1}.
\item
  \texttt{bwc\_impute1} \newline Scaling of calculated bandwidth, or if
  \texttt{explicit\_bandwidth\ =\ TRUE} used as the bandwidth. It is
  used in the imputation step for \(\hat{m}_1(\beta_1^T x)\). The
  default value is \texttt{1.25}.
\item
  \texttt{bwc\_dim\_red0}~\\
  Scaling of calculated bandwidth, or if
  \texttt{explicit\_bandwidth\ =\ TRUE} used as the bandwidth. It is
  used in the dimension reduction step for \(\hat{m}_0(\beta_0^T x)\).
  The default value is \texttt{1}.
\item
  \texttt{bwc\_impute0} \newline Scaling of calculated bandwidth, or if
  \texttt{explicit\_bandwidth\ =\ TRUE} used as the bandwidth. It is
  used in the imputation step for \(\hat{m}_0(\beta_0^T x)\). The
  default value is \texttt{1.25}.
\item
  \texttt{gauss\_cutoff}~\\
  The cutoff value for Gaussian kernel. The default value is
  \texttt{1e-3}.
\item
  \texttt{penalty} \newline Penalty for the optimizer if local linear
  regression fails. Added to the function value in solver as
  \texttt{penalty}\^{}(n - \texttt{n\_before\_pen}), where n is the
  number of times local linear regression fails. The default value is
  \texttt{10}.
\item
  \texttt{n\_before\_pen}~\\
  The number of acceptable local linear regression failures during the
  estimation of \(\beta_0\) and \(\beta_1\) phase. The default value is
  \texttt{5}.
\item
  \texttt{to\_extrapolate}~\\
  Specifies whether to extrapolate or not. Since in
  \(\hat{m}_0(\beta_0^T x)\) and \(\hat{m}_1(\beta_1^T x)\) estimates in
  terms of \(\beta_0^Tx\) and \(\beta_1^Tx\), local linear regression at
  the boundaries of \(\beta_0^TX\), and \(\beta_1^TX\), can be very
  volatile, it is recommended to use extrapolation on those points
  instead of local linear regression. The default value is
  \texttt{TRUE}.
\item
  \texttt{extrapolation\_basis} \newline The number of data points to
  base extrapolation on. Extrapolation at border points can be done
  based on a different number of neighborhood points.
  \texttt{extrapolation\_basis} is how many neighborhood points are
  used. The default value is \texttt{5}.
\item
  \texttt{to\_truncate} \newline Specifies whether to truncate
  \(\hat{m}_0(\beta_0^T x)\) and \(\hat{m}_1(\beta_1^T x)\) or not.
  After estimating \(\hat{m}_0(\beta_0^T x)\) and
  \(\hat{m}_1(\beta_1^T x)\), if they are outside the range of observed
  outputs, they are replaced with the minimum and maximum observed
  outputs. The default value is \texttt{TRUE}.
\item
  \texttt{n\_threads}~\\
  Sets the number of threads for parallel computing. Set to 1 serial. If
  \texttt{n\_threads} exceeds the maximum number of threads, sets
  \texttt{n\_threads} to max\_threads - 1. To use max\_threads, set to
  \texttt{n\_threads} to max\_threads of system. The default value is
  \texttt{1}.
\item
  \texttt{verbose} \newline Specifies if the program should print output
  while running. The default value is \texttt{TRUE}.
\item
  \ldots{}\\
  Additional parameters passed to \texttt{optim} or \texttt{cobyla}.
\end{itemize}

Output values include IMP and IMP2 estimators of ATE:

\begin{Shaded}
\begin{Highlighting}[]
\NormalTok{imp}\OperatorTok{$}\NormalTok{ate}
\CommentTok{#> [1] 2.054309}

\NormalTok{imp}\OperatorTok{$}\NormalTok{ate2}
\CommentTok{#> [1] 2.074988}
\end{Highlighting}
\end{Shaded}

Furthermore, the imputed values computed for each individual,
\(\widehat{m}_1(\hat{\beta}_1^T x_i)\) and
\(\widehat{m}_0(\hat{\beta}_0^T x_i)\), are stored in
\texttt{imp\$m1\$m} and \texttt{imp\$m0\$m}, and derivatives in
\texttt{imp\$m1\$dm} and \texttt{imp\$m0\$dm}. Finally, \(\hat\beta_0\)
and \(\hat\beta_1\) are also available as output:

\begin{Shaded}
\begin{Highlighting}[]
\NormalTok{imp}\OperatorTok{$}\NormalTok{beta1_hat}
\CommentTok{#>            [,1]}
\CommentTok{#> [1,]  1.0000000}
\CommentTok{#> [2,] -0.9867943}
\CommentTok{#> [3,]  0.9828903}
\CommentTok{#> [4,] -2.0523455}
\CommentTok{#> [5,] -1.3962939}
\CommentTok{#> [6,]  0.5248148}

\NormalTok{imp}\OperatorTok{$}\NormalTok{beta0_hat}
\CommentTok{#>             [,1]}
\CommentTok{#> [1,]  1.00000000}
\CommentTok{#> [2,]  1.02942153}
\CommentTok{#> [3,] -0.05913347}
\CommentTok{#> [4,] -0.11466994}
\CommentTok{#> [5,]  0.03370901}
\CommentTok{#> [6,]  0.04537406}
\end{Highlighting}
\end{Shaded}

\hypertarget{visualization}{%
\subsubsection{Visualization}\label{visualization}}

The visualization function provided for the IMP estimator is
\texttt{plot.imp}. It needs \texttt{ggplot2} to run. The function yields
three graphs. The first graph displays all observed responses and
estimated (imputed) potential outcomes against the index. The second
graph shows observed outcomes and estimated potential outcomes against
the estimated dimension reduction subspace \(\hat{\beta}_1^T x\). The
third graph is the same as the second one, but against the estimated
dimension reduction subspace \(\hat{\beta}_0^T x\).

The first plot (Figure 2) can be used, e.g., to check whether the range
and dispersion of the imputed values is similar to the distribution of
observed values for the treated and untreated groups. The second and
third plots (Figures 3 and 4) can be used, e.g., to inspect visually if
imputed values seem reasonable. The fitted values represent the
functions \(m_1\) and \(m_0\), respectively. The smoothness of the
estimated functions is related to the kernel bandwidths. In other words,
these graphs can also be used to help choosing appropriate bandwidths.
If the estimated function is too flexible and sensitive to minor errors
in observed values, then this is a sign that the selected bandwidths
were too small.

\begin{Shaded}
\begin{Highlighting}[]
\KeywordTok{plot}\NormalTok{(imp , }\DataTypeTok{covariates =}\NormalTok{ covariates, }\DataTypeTok{y=}\NormalTok{y, }\DataTypeTok{treated =}\NormalTok{ treated)}
\CommentTok{#> [1] "salam"}
\CommentTok{#> $pl_imp}
\end{Highlighting}
\end{Shaded}

\begin{figure}
\centering
\includegraphics{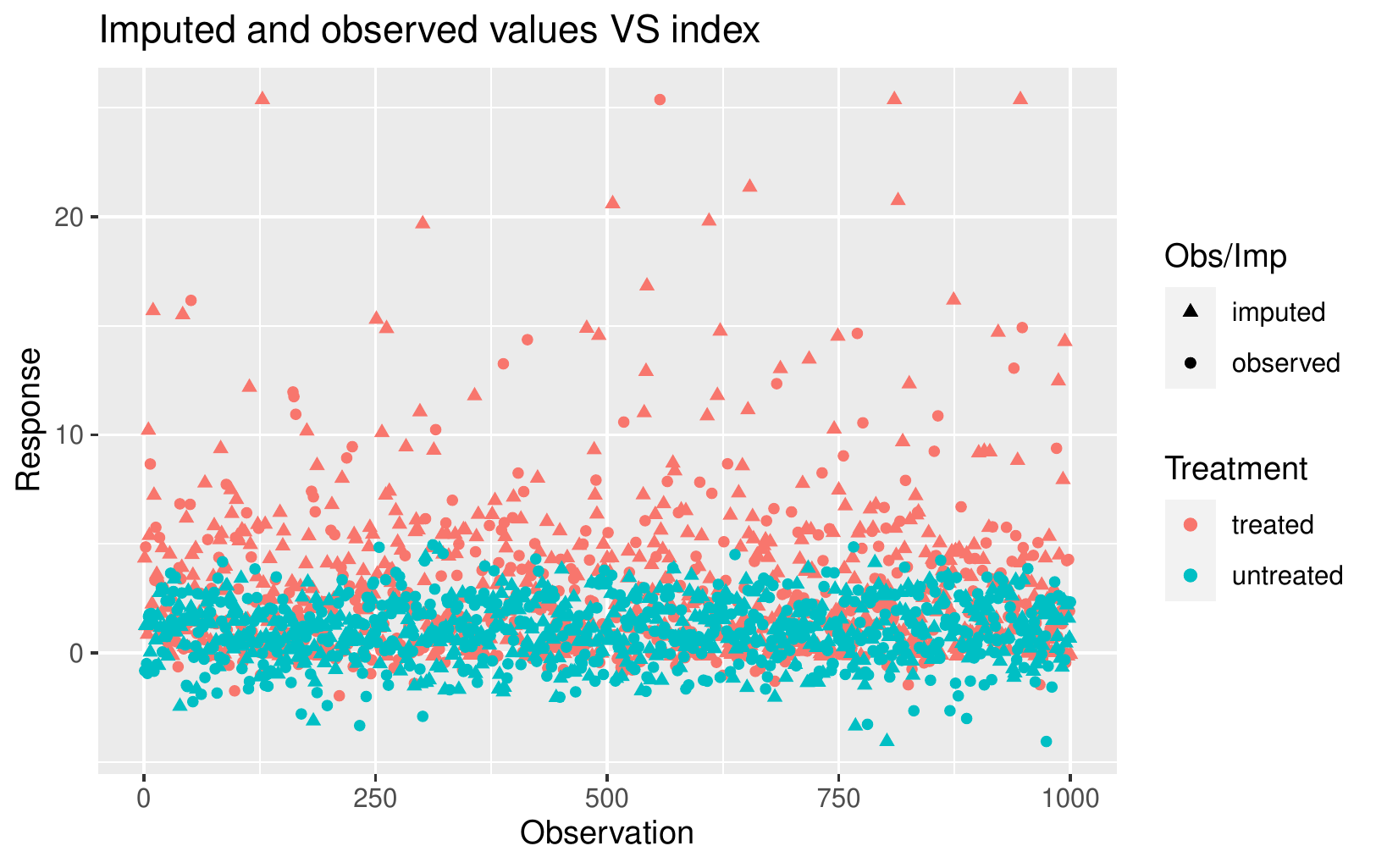}
\caption{The imputed and observed outcome values for each individual,
for the treated (red) and for the controls (blue).}
\end{figure}

\begin{verbatim}
#> 
#> $pl_m1
\end{verbatim}

\begin{figure}
\centering
\includegraphics{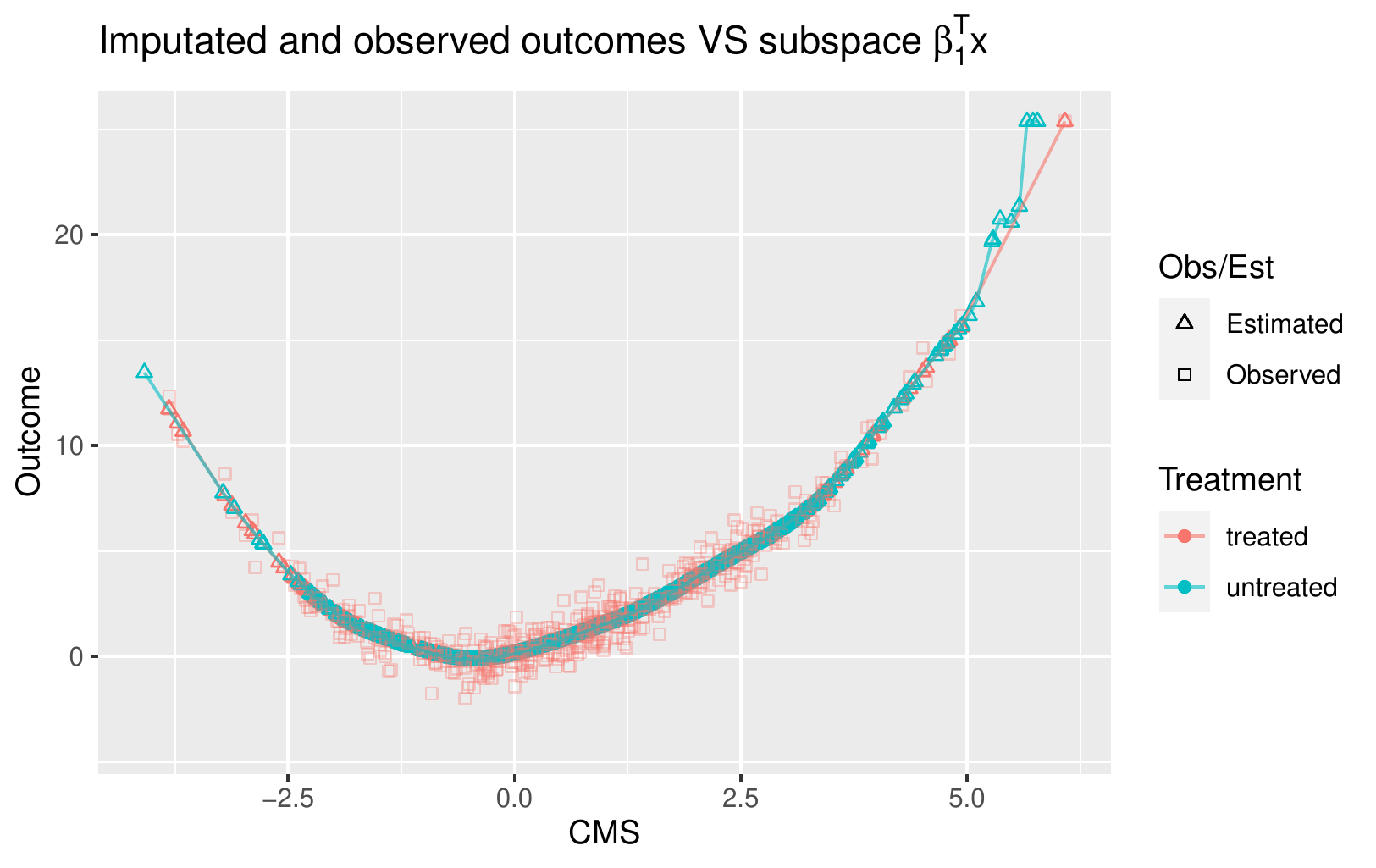}
\caption{The fitted (estimation of \(m_1\)) and observed outcome values
for the treated (red) and the predicited values for the controls (blue)
versus the estimated central mean subspace (CMS) for the treated.}
\end{figure}

\begin{verbatim}
#> 
#> $pl_m0
\end{verbatim}

\begin{figure}
\centering
\includegraphics{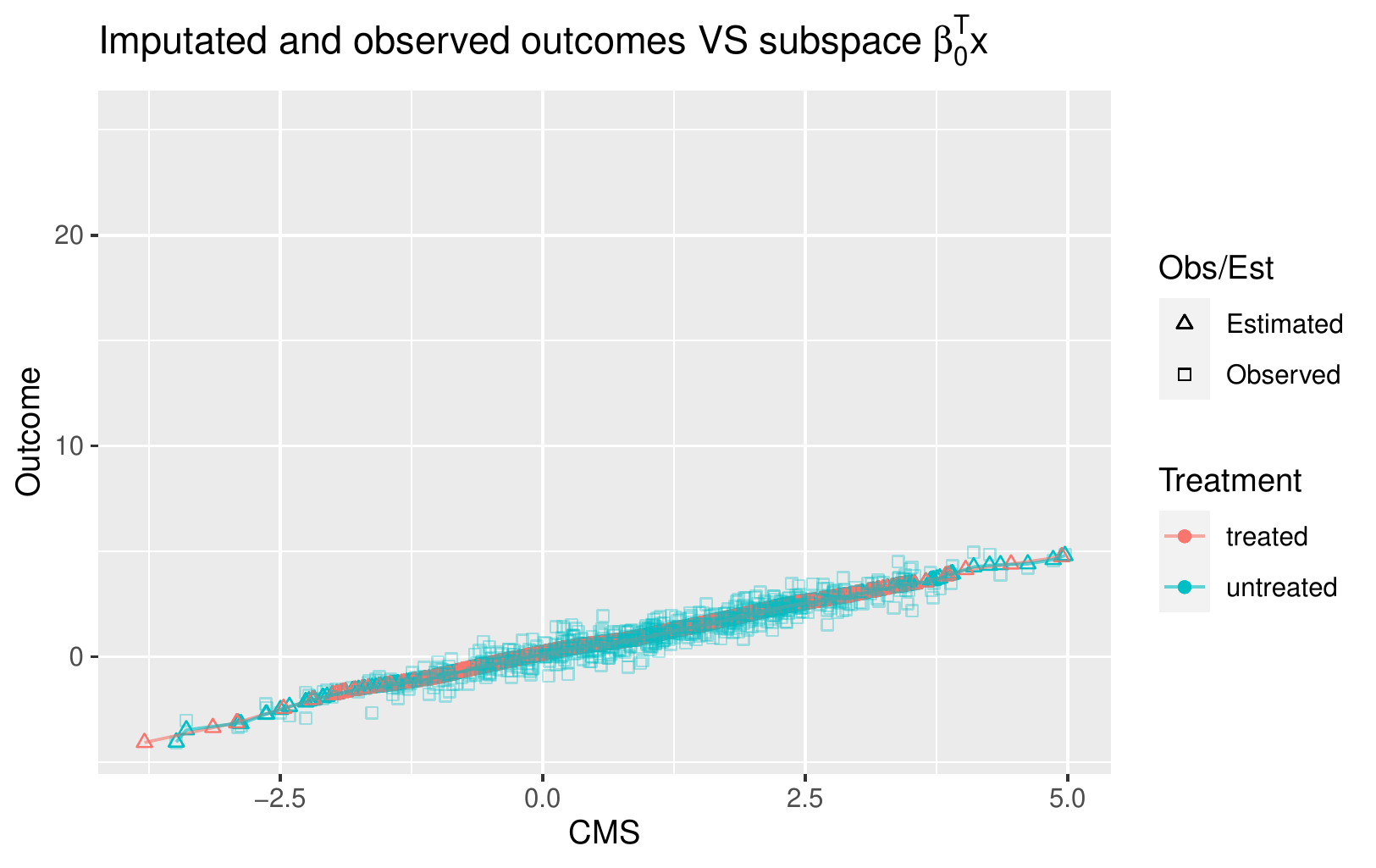}
\caption{The fitted (estimation of \(m_0\)) and observed values for the
controls (blue) and the predicted values for the treated (red) versus
the estimated central mean subspace for the controls.}
\end{figure}

The bandwidths computed are dependent on the standard deviation of the
projected covariates. The higher these variances, the larger the
bandwidths. Nevertheless, the distribution of data points in the
projection subspace can be non-uniform. In such settings, the bandwidth
may be too small for the low-density intervals, and the support of the
kernel function may not include neighboring data. If the kernel method
fails because of a low density of data points in some regions, the
package uses extrapolation and interpolation to impute missing values.
For a point inside the data region, a linear interpolation is performed
using two nearest points. For a point outside, extrapolation is used
using a user-specified number of nearest points. The package uses the
average slope between each pair of neighboring points to obtain a slope
for the linear extrapolation. The results of these choices for
interpolation and extrapolation may be checked visually using the above
plots.

\hypertarget{inverse-probability-weighting-ipw}{%
\subsection{Inverse Probability Weighting
(IPW)}\label{inverse-probability-weighting-ipw}}

\hypertarget{description-1}{%
\subsubsection{Description}\label{description-1}}

The IPW estimator of ATE weighs observed outcomes by the inverse of the
estimated propensity score. For details, see Section 3.2 in Ghosh, Ma,
and de~Luna (2021). The expected potential outcomes are estimated as
\[\widehat{E}(Y_1) = n^{-1} \sum_{i=1}^n t_iy_i/\hat{p}_i,\] and
\[\widehat{E}(Y_0) = n^{-1} \sum_{i=1}^n (1-t_i)y_i/(1- \hat{p}_i),\]
where
\(\hat p_i= e^{\hat{\eta}(\hat\alpha^Tx)}/(1+e^{\hat{\eta}(\hat\alpha^Tx)})\),
thereby yielding a corresponding estimator of ATE:
\[\widehat{D}_{IPW} = \widehat{E}(Y_1) - \widehat{E}(Y_0).\]

\hypertarget{example-and-code-1}{%
\subsubsection{Example and Code}\label{example-and-code-1}}

The function \texttt{ipw.ate} implements such an IPW estimator. Just as
the function \texttt{imp.ate} needed an initial guess for \(\beta_1\)
and \(\beta_0\), the function \texttt{ipw.ate} needs an initial guess
for \(\alpha\). Given how accurate the initial \(\alpha\) guess is, one
can choose the relevant optimizer. Selecting the optimizer function is
done by using the \texttt{solver\ =\ optim} or
\texttt{solver\ =\ cobyla} options. The option \texttt{method} can be
used to select different methods for the \texttt{optim} function.

In estimating the propensity score, the local linear regression
estimator may not be between 0 and 1. In such cases, we use linear
interpolation to obtain an estimate between 0 and 1. However, a large
number of such interpolations will reduce the accuracy of the propensity
score estimation. Therefore, controlling the number of interpolations
seems necessary. For this purpose, we consider this number as an
additive penalty in the optimization problem.

The user can add such a penalty to the optimization. The optimization
solver considers the sum of the loss function and the penalty term as
the new loss function. There are two variables \texttt{penalty} and
\texttt{n\_before\_pen} that should be assigned. The additive penalty
term is \texttt{penalty} to the power of the number of predicted
propensity scores falling outside of the interval {[}0,1{]}, minus
\texttt{n\_before\_pen}.

\[l_{New} = l + \textrm{penalty}^{(n - \textrm{n\_before\_pen})},\]

where \(n\) is the number of estimated propensity scores bigger than 1
or less than 0.

As for \texttt{imp.ate}, there is an option \texttt{n\_threads} allowing
parallelization, that one can use to decrease runtime for big datasets.
The maximum acceptable number can be found by calling the function
\texttt{parallel::detectCores()}. Only one thread is possible when using
Mac OS because OpenMP is not available on this operative system.

\begin{Shaded}
\begin{Highlighting}[]

\NormalTok{ipw <-}\StringTok{ }\KeywordTok{ipw.ate}\NormalTok{(}\DataTypeTok{x =}\NormalTok{ covariates,}
               \DataTypeTok{y =}\NormalTok{ y,}
               \DataTypeTok{treated =}\NormalTok{ treated,}
               \DataTypeTok{alpha_initial =}\NormalTok{ alpha)}
\end{Highlighting}
\end{Shaded}

The complete list of arguments is:

\begin{itemize}
\item
  \texttt{x} \newline Covariate matrix.
\item
  \texttt{y} \newline   Response vector.
\item
  \texttt{treated} \newline     A binary vector indicating treatment
  status.
\item
  \texttt{alpha\_initial} \newline   Initial guess for \(\alpha\)
\item
  \texttt{solver}~\\
  Specifies which solver is to be used. Current options are
  \texttt{optim} and \texttt{cobyla} (from \texttt{nloptr} package). The
  default value is \texttt{"optim"}.
\item
  \texttt{kernel}~\\
  Specifies which kernel function to be used, current options are:
  \texttt{"EPAN"}, \texttt{"QUARTIC"}, and \texttt{"GAUSSIAN"}. The
  default is \texttt{"EPAN"}.
\item
  \texttt{explicit\_bandwidth}~\\
  Specifies if \texttt{bandwidth\_scale} will be used as the bandwidth
  or if it will be calculated as \texttt{bandwidth\_scale} \(\times\)
  sd(\(\alpha^T x\)) \(\times\) \(n^{1/5}\). The default value is
  \texttt{FALSE}.
\item
  \texttt{recalc\_bandwidth}~\\
  Specifies whether the bandwidth should be recalculated after the
  estimation of \(\alpha\). If \texttt{explicit\_bandwidth} =
  \texttt{TRUE}, \texttt{recalc\_bandwidth} is not used, but if
  \texttt{explicit\_bandwidth} = \texttt{FALSE}, and
  \texttt{recalc\_bandwidth} = \texttt{TRUE}, bandwidth is recalculated
  at the beginning of the second step based on
  \texttt{bwc\_prop\_score}. If \texttt{recalc\_bandwidth} =
  \texttt{FALSE}, the first step bandwidth is used. The default value is
  \texttt{TRUE}.
\item
  \texttt{bwc\_dim\_red} \newline Scaling of calculated bandwidth, or if
  \texttt{explicit\_bandwidth\ =\ TRUE} used as the bandwidth. It is
  used in the dimension reduction step for \(\alpha^T x\). The default
  value is \texttt{1}.
\item
  \texttt{bwc\_prop\_score}~\\
  Scaling of calculated bandwidth, or if
  \texttt{explicit\_bandwidth\ =\ TRUE} used as the bandwidth. It is
  used for the estimation of the propensity score. The default value is
  \texttt{10}.
\item
  \texttt{gauss\_cutoff}~\\
  The cutoff value for Gaussian kernel. The default value is
  \texttt{1e-3}.
\item
  \texttt{penalty} \newline Penalty for the optimizer if a probability
  is outside (0, 1) during the estimation of \(\alpha\) phase. Added to
  the function value in solver as \texttt{penalty}\^{}(n -
  \texttt{n\_before\_pen}), where n is the number of probabilities
  outside (0, 1). The default value is \texttt{10}.
\item
  \texttt{n\_before\_pen}~\\
  The number of probabilities outside the range (0, 1) to accept during
  the estimation of \(\alpha\) phase. The default value is \texttt{1}.
\item
  \texttt{n\_threads} \newline The number of threads for parallel
  computing. Set it to 1 for serial. If \texttt{n\_threads} exceeds the
  maximum number of threads, \texttt{n\_threads} is set to max\_threads
  - 1. To use max\_threads, set \texttt{n\_threads} to max\_threads of
  the system. The default value is \texttt{1}.
\item
  \texttt{verbose} \newline Specifies if the program should print output
  while running. The default value if \texttt{TRUE}.
\item
  \ldots{}\\
  Additional parameters passed to \texttt{optim} or \texttt{cobyla}.
\end{itemize}

The IPW estimate of the average treatment effect is the main output of
the function:

\begin{Shaded}
\begin{Highlighting}[]
\NormalTok{ipw}\OperatorTok{$}\NormalTok{ate}
\CommentTok{#> [1] 2.174172}
\end{Highlighting}
\end{Shaded}

Other outputs are the fitted propensity scores and their first
derivatives.

\begin{Shaded}
\begin{Highlighting}[]
\NormalTok{P =}\StringTok{ }\NormalTok{ipw}\OperatorTok{$}\NormalTok{pr}
\NormalTok{dP =}\StringTok{ }\NormalTok{ipw}\OperatorTok{$}\NormalTok{d_pr}
\end{Highlighting}
\end{Shaded}

If the function is run by setting \texttt{explicit\_bandwidth} =
\texttt{TRUE}, then the bandwidths are computed based on the
coefficients \texttt{bwc\_dim\_red} and \texttt{bwc\_prop\_score}. The
bandwidths used are given as output:

\begin{Shaded}
\begin{Highlighting}[]
\NormalTok{ipw}\OperatorTok{$}\NormalTok{bw_dr}
\CommentTok{#> [1] 0.08774137}
\NormalTok{ipw}\OperatorTok{$}\NormalTok{bw_pr}
\CommentTok{#> [1] 0.8641581}
\end{Highlighting}
\end{Shaded}

as well as the estimated value of \(\alpha\):

\begin{Shaded}
\begin{Highlighting}[]
\NormalTok{ipw}\OperatorTok{$}\NormalTok{alpha_hat}
\CommentTok{#>             [,1]}
\CommentTok{#> [1,] -0.27000000}
\CommentTok{#> [2,]  0.19668843}
\CommentTok{#> [3,] -0.14153543}
\CommentTok{#> [4,]  0.03486261}
\CommentTok{#> [5,]  0.15451347}
\CommentTok{#> [6,] -0.08843496}
\end{Highlighting}
\end{Shaded}

\hypertarget{visualization-1}{%
\subsubsection{Visualization}\label{visualization-1}}

The \texttt{plot.ipw} function visualizes the fitted propensity scores
versus the estimated central mean subspace \(\hat{\alpha}^T x\),
together with the treatment status of each individual; see Figure 5.

\begin{Shaded}
\begin{Highlighting}[]
\KeywordTok{plot}\NormalTok{(ipw, }\DataTypeTok{treated =}\NormalTok{ treated, }\DataTypeTok{covariates =}\NormalTok{ covariates)}
\end{Highlighting}
\end{Shaded}

\begin{figure}
\centering
\includegraphics{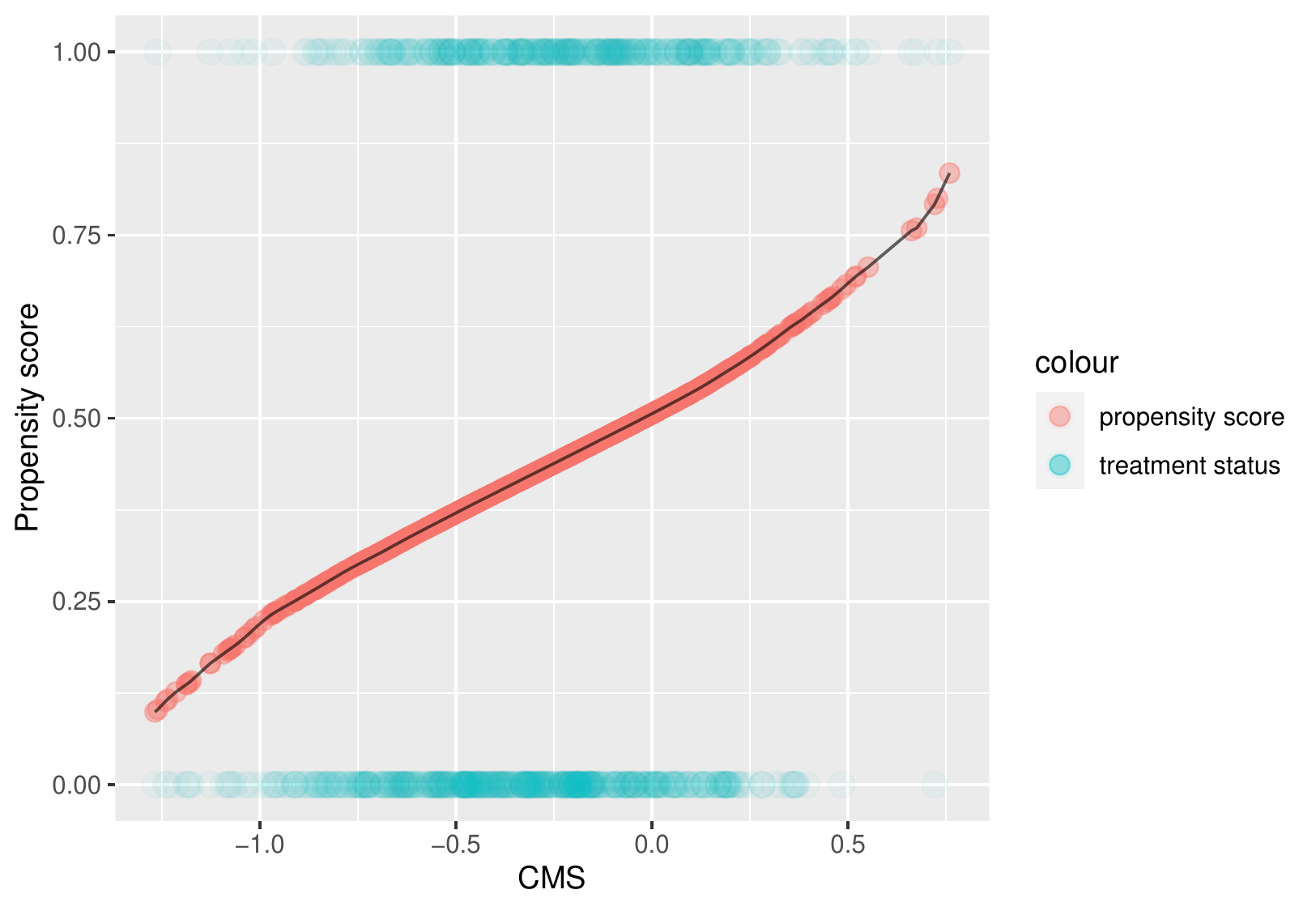}
\caption{The estimated propensity score (red) and the treatment status
(blue) versus the estimated central mean subspace.}
\end{figure}

\hypertarget{augmented-inverse-probability-weighting-aipw}{%
\subsection{Augmented Inverse Probability Weighting
(AIPW)}\label{augmented-inverse-probability-weighting-aipw}}

\hypertarget{description-2}{%
\subsubsection{Description}\label{description-2}}

The AIPW estimator combines fits from outcome and propensity score
models, and is thereby doubly robust, i.e.~consistent and asymptotically
normal when either the outcome models or the propensity score model is
correctly specified; see Ghosh, Ma, and de~Luna (2021). The estimator is
obtained as follows:

\[\widehat{E}(Y_1) = n^{-1} \sum_{i=1}^n \{ t_iy_i/\hat{p}_i + (1 - t_i/\hat{p}_i)\hat{m}_1(\hat{\beta}_1^T x_i)\},\]

\[\widehat{E}(Y_0) = n^{-1} \sum_{i=1}^n \{ (1-t_i)y_i/(1- \hat{p}_i) + [1- (1-t_i)/(1-\hat{p}_i)]\hat{m}_0(\hat{\beta}_0^T x_i) \},\]
and \[\widehat{D}_{AIPW} = \widehat{E}(Y_1) - \widehat{E}(Y_0). \]

\hypertarget{example-and-code-2}{%
\subsubsection{Example and Code}\label{example-and-code-2}}

The function \texttt{aipw.ate} that implements AIPW, asks for
\texttt{imp.ate} and \texttt{ipw.ate} objects as inputs as well as
observed values for outcome and treatment status.

\begin{Shaded}
\begin{Highlighting}[]
\NormalTok{ate_aipw <-}\StringTok{ }\KeywordTok{aipw.ate}\NormalTok{( }\DataTypeTok{y=}\NormalTok{y,}\DataTypeTok{treated =}\NormalTok{ treated,}\DataTypeTok{imp =}\NormalTok{ imp,}\DataTypeTok{ipw =}\NormalTok{ ipw)}
\end{Highlighting}
\end{Shaded}

Thus, inputs are:

\begin{itemize}
\item
  \texttt{y} \newline Observed response
\item
  \texttt{treated} \newline A binary vector indicating treatment
\item
  \texttt{imp} \newline The imp function output object from
  \texttt{imp.ate}
\item
  \texttt{ipw} \newline The ipw function output object from
  \texttt{ipw.ate}
\end{itemize}

Furthermore, the function returns the AIPW estimation of the average
treatment effect:

\begin{Shaded}
\begin{Highlighting}[]
\NormalTok{ate_aipw}
\CommentTok{#> [1] 2.025275}
\end{Highlighting}
\end{Shaded}

\hypertarget{improved-augmented-inverse-probability-weighting-aipw2}{%
\subsection{Improved Augmented Inverse Probability Weighting
(AIPW2)}\label{improved-augmented-inverse-probability-weighting-aipw2}}

\hypertarget{description-3}{%
\subsubsection{Description}\label{description-3}}

An improved version of the AIPW estimator is also implemented; see
details in Section 3.3 of Ghosh, Ma, and de~Luna (2021):

\[\widehat{E}(Y_1) = n^{-1} \sum_{i=1}^n \{ t_iy_i/\hat{p}_i + \hat{\gamma}_1(1 - t_i/\hat{p}_i)\hat{m}_1(\hat{\beta}_1^T x_i)\},\]

\[\widehat{E}(Y_0) = n^{-1} \sum_{i=1}^n \{ (1-t_i)y_i/(1- \hat{p}_i) + \hat{\gamma}_0[1- (1-t_i)/(1-\hat{p}_i)]\hat{m}_0(\hat{\beta}_0^T x_i) \},\]
and \[\widehat{D}_{IAIPW} = \widehat{E}(Y_1) - \widehat{E}(Y_0), \]
where \(\hat{\gamma}_0\), and \(\hat{\gamma}_1\) are:

\[\hat{\gamma}_1 = cov \{ \hat{m}_1 (\hat{\beta}_1^T x_i)t_i/\hat{p}_i , (1- t_i/\hat{p}_i)\hat{m}_1 (\hat{\beta}_1^T x_i)\}^{-1} \times  cov \{ t_i y_i/\hat{p}_i , (1- t_i/\hat{p}_i)\hat{m}_1 (\hat{\beta}_1^T x_i)\},\]

\[\hat{\gamma}_0 = cov \{(1-t_i)/(1- \hat{p}_i) \hat{m}_0 (\hat{\beta}_0^T x_i) ,  (t_i - \hat{p}_i)/(1-\hat{p}_i)\hat{m}_0 (\hat{\beta}_0^T x_i)\}^{-1} \times  cov \{ (1-t_i) y_i/(1-\hat{p}_i) ,  (t_i - \hat{p}_i)/(1-\hat{p}_i)\hat{m}_0 (\hat{\beta}_0^T x_i)\}.\]

\hypertarget{example-and-code-3}{%
\subsubsection{Example and Code}\label{example-and-code-3}}

The AIPW2 function has as same inputs as the AIPW function.

\begin{Shaded}
\begin{Highlighting}[]

\NormalTok{ate_iaipw <-}\StringTok{ }\KeywordTok{aipw2.ate}\NormalTok{(}\DataTypeTok{y=}\NormalTok{y,}\DataTypeTok{treated =}\NormalTok{ treated,}\DataTypeTok{imp =}\NormalTok{ imp,}\DataTypeTok{ipw=}\NormalTok{ipw)}
\end{Highlighting}
\end{Shaded}

\begin{itemize}
\item
  \texttt{y} \newline Observed response
\item
  \texttt{treated} \newline A binary vector indicating treatment
\item
  \texttt{imp} \newline The imp function output object from
  \texttt{imp.ate}
\item
  \texttt{ipw} \newline The ipw function output object from
  \texttt{ipw.ate}
\end{itemize}

As output, the ATE estimation is given:

\begin{Shaded}
\begin{Highlighting}[]
\NormalTok{ate_iaipw}
\CommentTok{#> [1] 2.025727}
\end{Highlighting}
\end{Shaded}

\hypertarget{variance-of-the-ate-estimators}{%
\section{Variance of the ATE
estimators}\label{variance-of-the-ate-estimators}}

In this section, we describe functions designed to calculate the
asymptotic variances of the estimators discussed above. These asymptotic
variances are obtained under regularity conditions, including:

\begin{itemize}
\item
  C1 The univariate \(m\)th order kernel function \(K(\cdot)\) is
  symmetric, Lipschitz continuous on its support \([-1, 1]\), which
  satisfies
  \[\int K(u)du=1, \int u^i K(u)du=0, 1\leq i \leq m-1, 0\neq \int u^m K(u)du < \infty. \]
\item
  C2 The bandwidths satisfy \(nh^{2m} \to 0\), \(nh^{2d} \to \infty\).
\end{itemize}

Theorems with asymptotic properties are given in Ghosh, Ma, and de~Luna
(2021), Section 3.4. For the estimators described above, we have
\(\sqrt{n}(\hat{D}_\text{est} - D) \xrightarrow[]{d} \text{N}(0,var_\text{est})\),
where \(var_\text{est}\), the asymptotic variance for each estimator,
\(var_{est}\) are implemented in the functions described below.

\hypertarget{imp}{%
\subsection{IMP}\label{imp}}

\hypertarget{description-4}{%
\subsubsection{Description}\label{description-4}}

The asymptotic variance of the IMP estimator is \begin{equation}
\begin{split}
var_{IMP} =& E(\{m_1(\beta_1^T x_i) -  m_0(\beta_0^T x_i) - E(Y_1) + E(Y_0)\}) \\
& + E[1 + e^{-\eta(\alpha^T X_i)} | \beta_1^T x_i] t_i \{y_{1i} - m_1(\beta_1^T x_i) \} \\
& - E[1 + e^{\eta(\alpha^T X_i)} | \beta_0^T x_i] (1-t_i) \{y_{0i} - m_0(\beta_0^T x_i) \} \\
& - E[(1 - p_i) vec \{ X_{Li} m^\prime_1 (\beta_1^T X_i)^T \}]^T B_1 t_i \{ y_{1i} - m_1(\beta_1^T x_i) \} \\
& \times vec [m^\prime_1 (\beta_1^T x_i) \otimes \{ x_{Li} - E(X_{Li} | \beta_1^T x_i) \}] \\
& + E[p_i vec \{ X_{Li} m^\prime_0 (\beta_0^T X_i)^T \}]^T B_0 (1-t_i) \{ y_{0i} - m_0(\beta_0^T x_i) \} \\
& \times vec [m^\prime_0 (\beta_0^T x_i) \otimes \{ x_{Li} - E(X_{Li} | \beta_0^T x_i) \}] )^{2},
\end{split}
\end{equation} and for IMP2 \begin{equation}
\begin{split}
v_{IMP2} =& E(\{m_1(\beta_1^T x_i) -  m_0(\beta_0^T x_i) - E(Y_1) + E(Y_0)\}) \\
& + E[p_i^{-1}| \beta_1^T x_i] t_i \{y_{1i} - m_1(\beta_1^T x_i) \} \\
& - E[(1 - p_i)^{-1} | \beta_0^T x_i] (1-t_i) \{y_{0i} - m_0(\beta_0^T x_i) \} \\
& - E[vec \{ X_{Li} m^\prime_1 (\beta_1^T X_i)^T \}]^T B_1 t_i \{ y_{1i} - m_1(\beta_1^T x_i) \} \\
& \times vec [m^\prime_1 (\beta_1^T x_i) \otimes \{ x_{Li} - E(X_{Li} | \beta_1^T x_i) \}] \\
& + E[vec \{ X_{Li} m^\prime_0 (\beta_0^T X_i)^T \}]^T B_0 (1-t_i) \{ y_{0i} - m_0(\beta_0^T x_i) \} \\
& \times vec [m^\prime_0 (\beta_0^T x_i) \otimes \{ x_{Li} - E(X_{Li} | \beta_0^T x_i) \}] )^{2},
\end{split}
\end{equation}

where \(B_0\), and \(B_1\) are

\[ B_0 = \Bigg\{ E\Bigg( \frac{ \partial \text{vec}[(1-T_i)\{ Y_{0i} - m_0(\beta_0^T X_i) \} m_0^\prime (\beta_0^T X_i) \otimes \{ X_{Li} - E(X_{Li}| \beta_0^T X_i) \} ] }{ \partial \text{vecl}(\beta_0)^\text{T} }\Bigg) \Bigg\}^{-1}, \]

\[ B_1 = \Bigg\{ E\Bigg( \frac{ \partial \text{vec}[T_i\{ Y_{1i} - m_1(\beta_1^T X_i) \} m_1^\prime (\beta_1^T X_i) \otimes \{ X_{Li} - E(X_{Li}| \beta_1^T X_i) \} ] }{ \partial \text{vecl}(\beta_1)^\text{T} }\Bigg) \Bigg\}^{-1} .\]

Here \(vecl[M]\) is the concatenation of the lower \((p-d) \times d\)
block of a \(p \times d\) matrix \(M\), and \(vec[M]\) is a vector made
by concatenation of a matrix \(M\).

\hypertarget{example-and-code-4}{%
\subsubsection{Example and Code}\label{example-and-code-4}}

The functions \texttt{imp.var} and \texttt{imp2.var} implement the above
variance estimators. These functions require \texttt{imp.ate} and
\texttt{ipw.ate} objects as input. The above variances require an
estimation of \(E(X_{Li} | \beta_t^T x_i)\). As before, the kernel
regression is used, so the variance functions require bandwidths as
input. A suggestion is to use the bandwidths used by \texttt{imp.ate}.
In this case, \texttt{explicit\_bandwidth} must be set to \texttt{TRUE}
(default value) in \texttt{imp.var} and \texttt{imp2.var}.

\begin{Shaded}
\begin{Highlighting}[]
\NormalTok{imp <-}\StringTok{ }\KeywordTok{imp.ate}\NormalTok{(}\DataTypeTok{x =}\NormalTok{ covariates,}
               \DataTypeTok{y =}\NormalTok{ y,}
               \DataTypeTok{treated =}\NormalTok{ treated,}
               \DataTypeTok{beta_guess1 =}\NormalTok{ beta1,}
               \DataTypeTok{beta_guess0 =}\NormalTok{ beta0)}
\NormalTok{ipw <-}\StringTok{ }\KeywordTok{ipw.ate}\NormalTok{(}\DataTypeTok{x =}\NormalTok{ covariates,}
               \DataTypeTok{y =}\NormalTok{ y,}
               \DataTypeTok{treated =}\NormalTok{ treated,}
               \DataTypeTok{alpha_initial =}\NormalTok{ alpha)}
\end{Highlighting}
\end{Shaded}

\begin{Shaded}
\begin{Highlighting}[]
\NormalTok{vimp <-}\StringTok{ }\KeywordTok{imp.var}\NormalTok{(}\DataTypeTok{x =}\NormalTok{ covariates,}
                \DataTypeTok{y =}\NormalTok{ y,}
                \DataTypeTok{treated =}\NormalTok{  treated,}
                \DataTypeTok{imp =}\NormalTok{  imp,}
                \DataTypeTok{ipw =}\NormalTok{  ipw)}
\NormalTok{vimp2 <-}\StringTok{ }\KeywordTok{imp2.var}\NormalTok{(}\DataTypeTok{x =}\NormalTok{ covariates,}
                  \DataTypeTok{y =}\NormalTok{ y,}
                  \DataTypeTok{treated =}\NormalTok{  treated,}
                  \DataTypeTok{imp =}\NormalTok{  imp,}
                  \DataTypeTok{ipw =}\NormalTok{  ipw)}
\end{Highlighting}
\end{Shaded}

\begin{itemize}
\item
  \texttt{x} \newline Covariate matrix
\item
  \texttt{y} \newline Response vector
\item
  \texttt{treated} \newline A binary vector indicating treatment
\item
  \texttt{imp} \newline The imp function output object from
  \texttt{imp.ate}
\item
  \texttt{ipw} \newline The ipw function output object from
  \texttt{ipw.ate}
\item
  \texttt{bandwidth\_scale1}~\\
  Scaling of the calculated bandwidth, or in case of
  \texttt{explicit\_bandwidth\ =\ TRUE}, the actual bandwidth for the
  estimation of \(E(\cdot|\beta_1^T X)\). The default value is
  \texttt{imp\$bw1}. If this default value is used, one should use the
  default value \texttt{TRUE} for \texttt{explicit\_bandwidth}.
\item
  \texttt{bandwidth\_scale0}~\\
  Scaling of the calculated bandwidth, or in case of
  \texttt{explicit\_bandwidth\ =\ TRUE}, the actual bandwidth for the
  estimation of \(E(\cdot|\beta_0^T X)\). The default value is
  \texttt{imp\$bw0}. If this default value is used, one should use the
  default value \texttt{TRUE} for \texttt{explicit\_bandwidth}.
\item
  \texttt{kernel}~\\
  Specifies which kernel function is to be used, current options are:
  \texttt{"EPAN"}(default), \texttt{"QUARTIC"}, and \texttt{"GAUSSIAN"}.
\item
  \texttt{explicit\_bandwidth}~\\
  Specifies if bandwidth\_scale will be used as the bandwidth or if it
  will be calculated as \texttt{bandwidth\_scale} \(\times\)
  sd(\(\beta_t^T x\)) \(\times\) \(n^{(1/5)}\). The default value is
  \texttt{TRUE}.
\item
  \texttt{gauss\_cutoff}~\\
  The cutoff value for Gaussian kernel. The default value is
  \texttt{1e-3}.
\end{itemize}

The output is an estimate of the asymptotic variance that can be used to
construct confidence intervals based on the asymptotic normal
distribution.

\begin{Shaded}
\begin{Highlighting}[]
\NormalTok{vimp}
\CommentTok{#> [1] 0.01671471}
\NormalTok{vimp2}
\CommentTok{#> [1] 0.0166378}
\end{Highlighting}
\end{Shaded}

\hypertarget{ipw}{%
\subsection{IPW}\label{ipw}}

\hypertarget{description-5}{%
\subsubsection{Description}\label{description-5}}

The asymptotic variance for the IPW estimator is:

\begin{equation}
\begin{split}
var_{IPW} =& E(\{ t_iy_{1i}/p_i - (1-t_i)y_{0i}/(1-p_i) - E(Y_1) + E(Y_0)\}) \\
& + (1 - t_i/p_i)E[m_1(\beta_1^T X_i) | \alpha^T x_i]  \\
& - (t_i - p_i)/(1 - p_i)E[m_0(\beta_0^T X_i) | \alpha^T x_i]  \\
& + (E[(m_{1i}(\beta_1^T X_i) (1-p_i) + m_{0i}(\beta_0^T X_i)p_i )vec \{ X_{Li} \eta^\prime (\alpha^T X_i)^T \}])^T B (t_i - p_i)  \\
& \times vec [ \{ x_{Li} - E(X_{Li} | \alpha^T x_i) \}\eta^\prime(\alpha^T x_i)^T  ] )^{2},
\end{split}
\end{equation}

where:
\[ B = \Bigg\{ E\Bigg( \frac{ \partial \text{vec}[  \{ X_{Li} - E(X_{Li}| \alpha^T X_i) \}(T_i - p_i) \eta^\prime (\alpha^T X_i)^T  ] }{ \partial \text{vecl}(\alpha)^\text{T} }\Bigg) \Bigg\}^{-1}. \]

\hypertarget{example-and-code-5}{%
\subsubsection{Example and Code}\label{example-and-code-5}}

In calculating this variance, \(E(X_{Li} | \alpha^T x_i)\) needs to be
estimated. Similar to the estimation of \(E(X_{Li} | \beta^T x_i)\) in
the functions \texttt{imp.var} and \texttt{imp2.var}, a kernel
regression is used to estimate \(E(X_{Li} | \alpha^T x_i)\) here.
Therefore, \texttt{ipw.var} needs a bandwidth as input too. The function
uses the bandwidth that comes from the output of \texttt{ipw.ate} as
long as the default values are used for \texttt{explicit\_bandwidth} and
\texttt{bandwidth\_scale}, i.e.~\texttt{explicit\_bandwidth\ =\ TRUE}
and \texttt{bandwidth\_scale\ =\ ipw\$bw\_dr}. The derivative in \(B\)
is difficult to find analytically, and numerical derivation is
performed. The argument \texttt{num\_deriv\_h} determines the accuracy
of the numerical derivation.

\begin{Shaded}
\begin{Highlighting}[]
\NormalTok{vipw <-}\StringTok{ }\KeywordTok{ipw.var}\NormalTok{(}\DataTypeTok{x =}\NormalTok{ covariates,}
                \DataTypeTok{y =}\NormalTok{ y,}
                \DataTypeTok{treated =}\NormalTok{ treated,}
                \DataTypeTok{imp =}\NormalTok{ imp,}
                \DataTypeTok{ipw =}\NormalTok{ ipw)}
\end{Highlighting}
\end{Shaded}

\begin{itemize}
\item
  \texttt{x} \newline   Covariate matrix
\item
  \texttt{y} \newline Response vector
\item
  \texttt{treated} \newline A binary vector indicating treatment
\item
  \texttt{imp} \newline imp output object from \texttt{imp.ate}
\item
  \texttt{ipw} \newline ipw output object from \texttt{ipw.ate}
\item
  \texttt{bandwidth\_scale} \newline Scaling of the calculated
  bandwidth, or in case of \texttt{explicit\_bandwidth\ =\ TRUE}, the
  actual bandwidth for the estimation of \(E(\cdot|\alpha^T X)\). The
  default value is \texttt{ipw\$bw\_dr}. If this default value is used,
  one should use the default value \texttt{TRUE} for
  \texttt{explicit\_bandwidth}.
\item
  \texttt{kernel}~\\
  Specifies which kernel function is to be used, current options are:
  \texttt{"EPAN"}(default), \texttt{"QUARTIC"}, and \texttt{"GAUSSIAN"}.
\item
  \texttt{explicit\_bandwidth}~\\
  Specifies if bandwidth\_scale will be used as the bandwidth or if it
  will be calculated as \texttt{bandwidth\_scale} \(\times\)
  sd(\(\alpha^T x\)) \(\times\) \(n^{(1/5)}.\) The default value is
  \texttt{TRUE}.
\item
  \texttt{gauss\_cutoff}~\\
  The cutoff value for Gaussian kernel. The default value is
  \texttt{1e-3}.
\item
  \texttt{num\_deriv\_h} \newline Step size of numerical derivative. The
  default value is \texttt{1e-8}.
\item
  \texttt{verbose} \newline Specifies if the program should print output
  while running. The default value if \texttt{FALSE}.
\end{itemize}

The output is an estimate of the asymptotic variance.

\begin{Shaded}
\begin{Highlighting}[]
\NormalTok{vipw}
\CommentTok{#> [1] 0.03044004}
\end{Highlighting}
\end{Shaded}

\hypertarget{aipw-and-iaipw}{%
\subsection{AIPW and IAIPW}\label{aipw-and-iaipw}}

\hypertarget{description-6}{%
\subsubsection{Description}\label{description-6}}

The asymptotic variances of the AIPW and the IAIPW estimators are equal.
One can obtain the variance by calling \texttt{aipw.var} and report it
for both estimators. \begin{equation}
\begin{split}
var_{IAIPW} = v_{AIPW} =& E(\{m_1(\beta_1^T x_i) -  m_0(\beta_0^T x_i) - E(Y_1) + E(Y_0)\}) \\
& + [1 + e^{-\eta(\alpha^T X_i)}] t_i \{y_{1i} - m_1(\beta_1^T x_i) \} \\
& - [1 + e^{\eta(\alpha^T X_i)} ] (1-t_i) \{y_{0i} - m_0(\beta_0^T x_i) \} \\
& - C_1 B_1 t_i \{ y_{1i} - m_1(\beta_1^T x_i) \} vec [m^\prime_1 (\beta_1^T x_i) \otimes \{ x_{Li} - E(X_{Li} | \beta_1^T x_i) \}] \\
& + C_0 B_0 (1-t_i) \{ y_{0i} - m_0(\beta_0^T x_i) \} vec [m^\prime_0 (\beta_0^T x_i) \otimes \{ x_{Li} - E(X_{Li} | \beta_0^T x_i) \}] \\
& + D_1 B (t_i - p_i)  vec [ \{ x_{Li} - E(X_{Li} | \alpha^T x_i) \}\eta^\prime(\alpha^T x_i)^T  ] \\
& + D_0 B (t_i - p_i)  vec [ \{ x_{Li} - E(X_{Li} | \alpha^T x_i) \}\eta^\prime(\alpha^T x_i)^T  ])^{2},
\end{split}
\end{equation}

where \(C_1\), \(C_0\), \(D_1\), and \(D_0\) are defined as:

\[C_1 = E[\{ \partial m_1(\beta_1^T X_i)/ \partial \text{vecl}(\beta_1)^T \}(1 - T_i/p_i)]\]
\[C_0 = E[\{ \partial m_0(\beta_0^T X_i)/ \partial \text{vecl}(\beta_0)^T \}(1 - (1 - T_i)/(1 - p_i))]\]
\[D_1 = E[\{Y_{1i} - m_1(\beta_1^T X_i) \} T_i e^{-\eta (\alpha^T X_i)} \text{vec}\{ X_{Li} \eta^\prime (\alpha^T X_i)^T \} ]\]
\[D_0 = E[\{Y_{0i} - m_0(\beta_0^T X_i) \} (1-T_i) e^{\eta (\alpha^T X_i)} \text{vec}\{ X_{Li} \eta^\prime (\alpha^T X_i)^T \} ].\]

\hypertarget{example-and-code-6}{%
\subsubsection{Example and Code}\label{example-and-code-6}}

Estimation of the variance of the AIPW estimator requires both
estimation of \(E(X_{Li} | \alpha^T x_i)\) and
\(E(X_{Li} | \beta_t^T x_i)\). As before, these estimates are performed
by kernel regression. Therefore, the function needs input values for the
bandwidths. The default is set to take the bandwidths used in the
\texttt{imp.ate} and \texttt{ipw.ate} functions.

\begin{Shaded}
\begin{Highlighting}[]
\NormalTok{vaipw =}\StringTok{ }\KeywordTok{aipw.var}\NormalTok{(}\DataTypeTok{x =}\NormalTok{ covariates,}
                 \DataTypeTok{y =}\NormalTok{ y,}
                 \DataTypeTok{treated =}\NormalTok{  treated,}
                 \DataTypeTok{imp =}\NormalTok{  imp,}
                 \DataTypeTok{ipw =}\NormalTok{  ipw)}
\end{Highlighting}
\end{Shaded}

\begin{itemize}
\item
  \texttt{x} \newline Covariate matrix
\item
  \texttt{y} \newline Response vector
\item
  \texttt{treated} \newline A binary vector indicating treatment
\item
  \texttt{imp} \newline imp output object from \texttt{imp.ate}
\item
  \texttt{ipw} \newline ipw output object from \texttt{ipw.ate}
\item
  \texttt{bandwidth\_scale1}~\\
  Scaling of the calculated bandwidth, or in case of
  \texttt{explicit\_bandwidth\ =\ TRUE}, the actual bandwidth for the
  estimation of \(E(\cdot|\beta_1^T X)\). The default value is
  \texttt{imp\$bw1}. If this default value is used, one should use the
  default value \texttt{TRUE} for \texttt{explicit\_bandwidth}.
\item
  \texttt{bandwidth\_scale0}~\\
  Scaling of the calculated bandwidth, or in case of
  \texttt{explicit\_bandwidth\ =\ TRUE}, the actual bandwidth for the
  estimation of \(E(\cdot|\beta_0^T X)\). The default value is
  \texttt{imp\$bw0}. If this default value is used, one should use the
  default value \texttt{TRUE} for \texttt{explicit\_bandwidth}.
\item
  \texttt{bandwidth\_scale\_pr}~\\
  Scaling of the calculated bandwidth, or in case of
  \texttt{explicit\_bandwidth\ =\ TRUE}, the actual bandwidth for the
  estimation of \(E(\cdot|\alpha^T X)\). The default value is
  \texttt{ipw\$bw\_dr}. If this default value is used, one should use
  the default value \texttt{TRUE} for \texttt{explicit\_bandwidth}.
\item
  \texttt{kernel}~\\
  Specifies which kernel function is to be used, current options are:
  \texttt{"EPAN"}(default), \texttt{"QUARTIC"}, and \texttt{"GAUSSIAN"}.
\item
  \texttt{explicit\_bandwidth}~\\
  Specifies if \texttt{bandwidth\_scale} will be used as the bandwidth
  or if it will be calculated as \texttt{bandwidth\_scale} \(\times\)
  sd(\(\beta^T x\)) \(\times\) \(n^{(1/5)}.\) The default value is
  \texttt{TRUE}.
\item
  \texttt{gauss\_cutoff}~\\
  Cutoff value for Gaussian kernel. The default value is \texttt{1e-3}.
\item
  \texttt{num\_deriv\_h} \newline Step size of numerical derivative. The
  default value is \texttt{1e-6}.
\item
  \texttt{verbose} \newline Specifies if the program should print output
  while running. The default value is \texttt{FALSE}.
\end{itemize}

The output is an estimate of the asymptotic variance that can be used to
construct confidence intervals based on the asymptotic normal
distribution.

\begin{Shaded}
\begin{Highlighting}[]
\NormalTok{vaipw}
\CommentTok{#> [1] 0.01640422}
\end{Highlighting}
\end{Shaded}

\hypertarget{general-function}{%
\section{General function}\label{general-function}}

The comprehensive function \texttt{inf.ate} executes all the functions
listed above, and the output contains all the outputs from these
functions. The input variables for the general function includes the
combination of the inputs for \texttt{ate.imp} and \texttt{ate.ipw}. The
functions \texttt{aipw.ate} and \texttt{aipw2.ate}, and the variance
functions use their default input values.

\begin{Shaded}
\begin{Highlighting}[]
\NormalTok{inf =}\StringTok{ }\KeywordTok{inf.ate}\NormalTok{(}\DataTypeTok{x =}\NormalTok{ covariates,}
               \DataTypeTok{y =}\NormalTok{ y,}
               \DataTypeTok{treated =}\NormalTok{ treated,}
               \DataTypeTok{beta_guess1 =}\NormalTok{ beta1,}
               \DataTypeTok{beta_guess0 =}\NormalTok{ beta0,}
               \DataTypeTok{alpha_initial =}\NormalTok{ alpha)}
\end{Highlighting}
\end{Shaded}

\begin{itemize}
\item
  \texttt{x} \newline Covariate matrix
\item
  \texttt{y} \newline Response vector
\item
  \texttt{treated1} \newline A binary vector indicating treatment status
\item
  \texttt{beta\_guess1} \newline Initial guess for \(\beta_1\)
\item
  \texttt{beta\_guess0} \newline Initial guess for \(\beta_0\)
\item
  \texttt{alpha\_initial} \newline Initial guess for \(\alpha\)
\item
  \texttt{imp.solver}~\\
  Specifies which solver is to be used for the imputation estimator.
  Current options are \texttt{optim} and \texttt{cobyla} (from
  \texttt{nloptr} package). The default value is \texttt{"optim"}.
\item
  \texttt{imp.kernel}~\\
  Specifies which kernel function is to be used for the imputation
  estimator, current options are: \texttt{"EPAN"}, \texttt{"QUARTIC"},
  and \texttt{"GAUSSIAN"}. The default value is \texttt{"EPAN"}.
\item
  \texttt{imp.explicit\_bandwidth}~\\
  Specifies if bandwidth\_scale will be used as the bandwidth or if it
  will be calculated as \texttt{bandwidth\_scale} \(\times\)
  sd(\(\beta^T x\)) \(\times\) \(n^{(1/5)}\). The default value is
  \texttt{FALSE}.
\item
  \texttt{imp.recalc\_bandwidth}~\\
  Specifies whether the bandwidth should be recalculated after the first
  stage. If \texttt{explicit\_bandwidth} is \texttt{TRUE},
  \texttt{imp.recalc\_bandwidth} is not used, but if
  \texttt{explicit\_bandwidth} is \texttt{FALSE}, then if
  \texttt{recalc\_bandwidth} is \texttt{TRUE}, bandwidths are
  recalculated at the beginning of the second step based on
  \texttt{bwc\_impute0} and \texttt{bwc\_impute1}. If
  \texttt{recalc\_bandwidth} is \texttt{FALSE}, the first step
  bandwidths are used. The default value is \texttt{FALSE}.
\item
  \texttt{bwc\_dim\_red1} \newline Scaling of calculated bandwidth, or
  if \texttt{explicit\_bandwidth\ =\ TRUE} used as the bandwidth. It is
  used in the dimension reduction step for \(\hat{m}_1(\beta_1^T x)\).
  The default value is \texttt{1}.
\item
  \texttt{bwc\_impute1} \newline Scaling of calculated bandwidth, or if
  \texttt{explicit\_bandwidth\ =\ TRUE} used as the bandwidth. It is
  used in the imputation step for \(\hat{m}_1(\beta_1^T x)\). The
  default value is \texttt{1.25}.
\item
  \texttt{bwc\_dim\_red0}~\\
  Scaling of calculated bandwidth, or if
  \texttt{explicit\_bandwidth\ =\ TRUE} used as the bandwidth. It is
  used in the dimension reduction step for \(\hat{m}_0(\beta_0^T x)\).
  The default value is \texttt{1}.
\item
  \texttt{bwc\_impute0} \newline Scaling of calculated bandwidth, or if
  \texttt{explicit\_bandwidth\ =\ TRUE} used as the bandwidth. It is
  used in the imputation step for \(\hat{m}_0(\beta_0^T x)\). The
  default value is \texttt{1.25}.
\item
  \texttt{imp.gauss\_cutoff}~\\
  The cutoff value for Gaussian kernel. The default value is
  \texttt{1e-3}.
\item
  \texttt{imp.penalty} Penalty for the optimizer if local linear
  regression fails. Added to the function value in solver as
  \texttt{penalty}\^{}(n - \texttt{n\_before\_pen}), where n is the
  number of times local linear regression fails. The default value is
  \texttt{10}.
\item
  \texttt{imp.n\_before\_pen}~\\
  The number of acceptable local linear regression failures during the
  estimation of \(\beta_0\) and \(\beta_1\) phase. The default value is
  \texttt{5}.
\item
  \texttt{imp.to\_extrapolate}~\\
  Specifies whether to extrapolate or not. Since in
  \(\hat{m}_0(\beta_0^T x)\) and \(\hat{m}_1(\beta_1^T x)\) estimates in
  terms of \(\beta_0\) and \(\beta_1\), local linear regression at the
  boundaries of \(\beta_0^TX\), and \(\beta_1^TX\), can be very
  volatile, it is recommended to use extrapolation on those points
  instead of local linear regression. The default value is
  \texttt{TRUE}.
\item
  \texttt{imp.extrapolation\_basis}~\\
  The number of data points to base extrapolation on. Extrapolation at
  border points can be done based on a different number of neighborhood
  points. \texttt{extrapolation\_basis} is how many neighborhood points
  are used. The default value is \texttt{5}.
\item
  \texttt{imp.to\_truncate} \newline Specifies whether to truncate
  \(\hat{m}_0(\beta_0^T x)\) and \(\hat{m}_1(\beta_1^T x)\) or not.
  After estimating \(\hat{m}_0(\beta_0^T x)\) and
  \(\hat{m}_1(\beta_1^T x)\), if they are outside the range of observed
  outputs, they are replaced with the minimum and maximum observed
  outputs. The default value is \texttt{TRUE}.
\item
  \texttt{n\_threads} \newline Sets the number of threads for parallel
  computing. Set to 1 serial. If n\_threads exceeds the maximum number
  of threads, sets \texttt{n\_threads} to \texttt{max\_threads} - 1. To
  use \texttt{max\_threads}, set to \texttt{n\_threads} to
  \texttt{max\_threads} of system. The default value is \texttt{1}.
\item
  \texttt{imp.solver.options} \newline List of parameters passed to
  optim or cobyla. The default value is \texttt{NA}.
\item
  \texttt{ipw.solver}~\\
  Specifies which solver is to be used for the IPW estimator. Current
  options are \texttt{optim} and \texttt{cobyla} (from \texttt{nloptr}
  package). The default value is
  \texttt{\textquotesingle{}optim\textquotesingle{}}.
\item
  \texttt{ipw.kernel}~\\
  Specifies which kernel function to be used for the IPW estimator,
  current options are: \texttt{"EPAN"}, \texttt{"QUARTIC"}, and
  \texttt{"GAUSSIAN"}. The default is \texttt{"EPAN"}.
\item
  \texttt{ipw.explicit\_bandwidth}~\\
  Specifies if bandwidth\_scale will be used as the bandwidth or if it
  will be calculated as \texttt{bandwidth\_scale} \(\times\)
  sd(\(\alpha^T x\)) \(\times\) \(n^{1/5}\). The default value is
  \texttt{FALSE}.
\item
  \texttt{ipw.recalc\_bandwidth}~\\
  Specifies whether the bandwidth should be recalculated after the
  estimation of \(\alpha\) (the estimation of dimension reduction step).
  If \texttt{explicit\_bandwidth} is \texttt{TRUE},
  \texttt{ipw.recalc\_bandwidth} is not used, but if
  \texttt{explicit\_bandwidth} is \texttt{FALSE}, then if
  \texttt{recalc\_bandwidth} is \texttt{TRUE}, bandwidth is recalculated
  at the beginning of the second step based on
  \texttt{bwc\_prop\_score}. If \texttt{recalc\_bandwidth} is
  \texttt{FALSE}, the first step bandwidth is used. The default value is
  \texttt{TRUE}.
\item
  \texttt{bwc\_dim\_red} \newline Scaling of calculated bandwidth, or if
  \texttt{explicit\_bandwidth\ =\ TRUE} used as the bandwidth. It is
  used in the dimension reduction step for \(\alpha^T x\). The default
  value is \texttt{1}.
\item
  \texttt{bwc\_prop\_score}~\\
  Scaling of calculated bandwidth, or if
  \texttt{explicit\_bandwidth\ =\ TRUE} used as the bandwidth. It is
  used for the estimation of the propensity score. The default value is
  \texttt{10}.
\item
  \texttt{ipw.gauss\_cutoff}~\\
  The cutoff value for Gaussian kernel. The default value is
  \texttt{1e-3}.
\item
  \texttt{ipw.penalty} \newline Penalty for the optimizer if a
  probability is outside (0, 1) during the estimation of \(\alpha\)
  phase. Added to the function value in solver as \texttt{penalty}\^{}(n
  - \texttt{n\_before\_pen}), where n is the number of probabilities
  outside (0, 1). The default value is \texttt{10}.
\item
  \texttt{ipw.n\_before\_pen}~\\
  The number of probabilities outside the range (0, 1) to accept during
  the estimation of \(\alpha\) phase. The default value is \texttt{1}.
\item
  \texttt{ipw.solver.options} \newline List of parameters passed to
  optim or cobyla. The default value is \texttt{NA}.
\item
  \texttt{verbose} \newline Specifies if the program should print output
  while running. The default value is \texttt{TRUE}.
\end{itemize}

Output is a list containing:

\begin{itemize}
\item
  \texttt{imp} \newline The output obejct of the function
  \texttt{imp.ate}.
\item
  \texttt{ipw} \newline The output obejct of the function
  \texttt{ipw.ate}.
\item
  \texttt{aipw} \newline The AIPW estimation of the average treatment
  effect.
\item
  \texttt{aipw2} \newline The improved AIPW estimation of the average
  treatment effect.
\item
  \texttt{imp\_var} \newline The asymptotic variance of the IMP
  estimator.
\item
  \texttt{imp2\_var} \newline The asymptotic variance of the IMP2
  estimator.
\item
  \texttt{ipw\_var} \newline The asymptotic variance of the IPW
  estimator.
\item
  \texttt{aipw\_var} \newline The asymptotic variance for the AIPW and
  AIPW2 estimators.
\end{itemize}

\hypertarget{acknowledgements}{%
\section*{Acknowledgements}\label{acknowledgements}}

We are grateful to Filip Edström who participated in the development of the package described herein, and to Emma Persson for her comments that have improved this
vignette. We also acknowledge research grants from the Marianne and
Marcus Wallenberg Foundation and from the Swedish Research Council.

\hypertarget{declaration-of-interests}{%
\section*{Declaration of interests}\label{declaration-of-interests}}

The authors declare that they have no known competing financial
interests or personal relationships that could have appeared to
influence the work reported in this paper.

\hypertarget{references}{%
\section*{References}\label{references}}
\addcontentsline{toc}{section}{References}

\hypertarget{refs}{}
\leavevmode\hypertarget{ref-ghosh2018sufficient}{}%
Ghosh, Trinetri, Yanyuan Ma, and Xavier de~Luna. 2021. ``Sufficient
Dimension Reduction for Feasible and Robust Estimation of Average Causal
Effect.'' \emph{Statistica Sinica} 31.
\url{https://doi.org/10.5705/ss.202018.0416}.

\end{document}